# Ideology, institutions, and economic growth: panel evidence 1995–2022

*An empirical reassessment using V-Dem, Heritage, and WDI indices*

*Eduardo Koffmann Jopia.*

*Email: eakoffma@uc.cl*

*Patricia Galilea Aranda*

*Email: pga@uc.cl*

*School of Engineering, Pontifical Catholic University of Chile.*

**Abstract:**

*Does it matter whether a government is "left-wing" or "right-wing" for economic growth? Using a panel of 113 countries (1995–2022), we combine: (i) the economic ideology of the executive branch (V-Dem), (ii) the disaggregated institutional quality of the economic freedom index (Heritage), separating a core institutional block (HN: property rights, government integrity, judicial effectiveness) from a block of liberalization policies (HL: trade/financial openness, regulatory efficiency, size of government), and (iii) economic performance (GDP per capita PPP and its growth, World Bank). We estimate panel models with fixed effects by country and year, and standard errors grouped by country. We find that HN is strongly associated with higher income levels, while HL, on average globally, is not significant for income level once HN is controlled for, nor does it consistently predict short-term growth. Government ideology exhibits weak direct effects on growth, although it operates indirectly via HL (right-wing governments tend to have higher HL scores). Robustness tests with long five-year differences and country-specific trends reinforce that, in the long run, "institutions prevail over politics." We conclude with recommendations for developing countries: prioritize the institutional core (rule of law, anti-corruption, effective justice) as a basis for policies—whether left-wing or right-wing—to bear fruit in terms of growth.*

<u>*Keywords: economic growth; institutions; ideology; economic freedom; country panel; fixed effects.*</u>

**Abstract:**

*Do left- or right-wing governments drive growth? Using a panel of 113 countries (1995–2022) we combine: (i) executive's economic ideology (V-Dem), (ii) economic freedom (Heritage) split into a core institutional block (HN: property rights, government integrity, judicial effectiveness) and a market liberalization block (HL: trade, investment, financial openness; regulatory efficiency; government size), and (iii) economic performance (PPP GDP per capita and growth, World Bank). Fixed-effects panel models with clustered errors show that HN is strongly associated with higher income levels, whereas HL is, on average, insignificant for income once HN is controlled and does not*

*robustly predict short-run growth. Ideology's direct effect on growth is weak, though it operates indirectly via HL (right-leaning governments tend to have higher HL scores). Five-year long-difference and country-specific trend tests reinforce that "institutions rule" for long-run development. Policy advice for developing countries: build core institutions first so that policies—left or right—can deliver growth.*

## 1. Introduction

Public debate often presents a dichotomy between "market" and "state," with the right emphasizing open markets and less state intervention, and the left emphasizing more state intervention and redistribution. However, the literature and comparative experiences indicate that the key to economic progress is the strength of institutions: property protection, low corruption, and effective justice allow both social democratic and liberal programs to achieve sustainable results. In Latin America, where there was rule of law and checks and balances, policies worked; institutional weakening under ideological projects led to only temporary achievements.

This paper tests two hypotheses with recent data: (H1) strong economic institutions (core block of the Heritage index) are positively associated with per capita income; (H2) the economic ideology of the executive branch has a weak direct effect on growth, although it could act indirectly via reforms that affect economic liberalization. The main contribution is to break down the economic freedom index (Heritage Foundation) into an institutional block (HN) and a liberalization block (HL) and estimate their independent relationship with income and growth. The design uses only fixed effects by country and year, without additional macroeconomic controls, with the variation identified within each country over time; this simplifies interpretation and avoids collinearity.

The Heritage Index of Economic Freedom is often associated with higher per capita income (Heritage Foundation, 2022), but it includes both institutional quality (respect for property, absence of corruption) and market policies (openness, low tax burden). We suspect that not all sub-dimensions of "economic freedom" contribute equally to development; high scores may be due mainly to deep institutions. Separating these blocks helps to distinguish which part drives growth and which is merely ideological.

The article is organized as follows: section 2 reviews the literature on institutions, ideology, and growth; section 3 describes key data, sources, and variables; section 4 details the empirical design and models; section 5 presents main results and robustness tests; section 6 discusses findings and causality; section 7 offers recommendations for developing countries; section 8 mentions limitations and future lines of research.

## 2. Relationship to the literature

A wealth of theoretical and empirical literature has demonstrated that institutional quality is key to long-term economic development. The New Institutional Economics, with authors such as North (1990), argues that stable institutions reduce uncertainty and transaction costs, protect property rights, and encourage investment and innovation. Historically, North and Thomas (1973) and Jones (1981) already linked efficient institutions with the emergence of prosperous economies. Barro (1991) showed that the rule of law, political stability, and low conflict correlate with higher growth. Hall and Jones (1999) found that differences in "social infrastructure" explain much of the variation in productivity between countries. Knack and Keefer (1995) and Mauro (1995) added that better property protection and less corruption are associated with higher subsequent growth.

The study by Acemoglu, Johnson, and Robinson (2001) strengthened the causal thesis of institutions, linking the mortality of European settlers to current institutional quality: countries with inclusive institutions bequeathed by colonizers show higher per capita income. Engerman & Sokoloff (1997), La Porta et al. (1998), and Rodrik, Subramanian & Trebbi (2004) reinforce that "institutions are the fundamental cause of long-term growth." Countries without inclusive institutions find it difficult to achieve sustained development, while with strong institutions, various strategies can lead to prosperity.

There are debates about causality. Glaeser et al. (2004) suggest that economic growth may precede institutional improvements, and some institutional indicators may be endogenous. Examples such as South Korea and China show high growth under authoritarian regimes, investing in human capital and macroeconomic policies without full democratic institutions, which only improved later. However, the dominant view continues to support the importance of effective institutions. Acemoglu and Robinson (2012) warn that "developmental dictatorships" are exceptional and face risks of stagnation if they do not transition to inclusive institutions. In short, the evidence suggests that institutions such as the rule of law and competent bureaucracies drive growth, while extractive institutions slow it down.

Regarding the role of government ideology (left vs. right) in economic performance, traditional theory suggests that left-wing governments tend to favor higher spending and a larger state, while right-wing governments prioritize fiscal discipline and deregulation. Studies such as Imbeau, Pétry & Lamari (2001) and Potrafke (2017) find ideological differences in public spending, but little systematic divergence in growth or unemployment. Potrafke (2017) concludes that since the 1990s, partisan differences in macroeconomic variables have been attenuated by globalization and institutional constraints. Previous meta-analyses (Imbeau et al., 2001) show a moderate relationship between ideology and government size, with no robust impact on growth.

In developing countries, institutional quality outweighs ideology. Navarrete and Ritzen (2021), in 20 Latin American countries, found that institutional quality has a positive impact on GDP per capita growth, while ideological orientation has no significant direct effect. Even higher government spending only boosts growth if there are strong institutions, becoming negative in weak institutional environments. It is not declared ideology, but institutional quality and policy coherence that determine economic performance. Left-wing governments can succeed if they maintain legal certainty; right-wing governments fail if they allow corruption or weaken institutions. Thus, strong institutions moderate ideological swings and channel any political program toward sustainable results. This thesis of "institutions rule" is tested with recent data, clarifying which institutional aspects matter most and how they interact with the government's economic ideology.

### 3. Data and sources

Government ideology: The executive branch's economic ideological orientation is measured using the V-Dem index (Variable v2exlrgen_osp), which places the government on a left-right spectrum. This variable reflects the stance on the size of the state, economic intervention, redistributive policies, trade openness, and macroeconomic discipline by country and year. Values range from -3 (left, nationalizations, redistribution) to +3 (right, free market, minimal state role). The LR (Left–Right) variable indicates left-wing (negative values) and right-wing (positive values) governments. This measure covers only the economic-fiscal axis and not cultural or political regime issues.

Economic institutions: we use the Heritage Foundation's Index of Economic Freedom, which brings together 12 annual components since 1995, such as judicial quality and trade openness. We regroup these into two blocks:

- HN (Institutional Core): Property Rights, Government Integrity, and Judicial Effectiveness, reflecting the rule of law and basic institutional quality.
- HL (Market liberalization): The other nine subcomponents: trade freedom, foreign investment, financial freedom, regulatory efficiency, and government size. These indicate the degree of economic liberalization and less state intervention.

Each component ranges from 0 to 100. HN and HL are obtained by averaging the corresponding values. Separating these blocks allows us to see which part of the index is most closely associated with economic development.

Economic performance: We measure GDP per capita (constant 2017 PPP dollars) and its real annual growth, using the natural logarithm and its annual difference to capture growth rates.

Country sample and period: data is collected from 113 countries between 1995 and 2022, including most advanced and developing economies. According to the 2017–2019 average, the top tertile of HN had a per capita income 7.1 times that of the bottom tertile; for HL it was ~3.95 and for ideology only ~1.63. This indicates that core institutions show the largest development gap. HN varies more between countries than over time; HL and ideology show more temporal variability.

Methodological note: the models do not include classic macroeconomic controls, only fixed effects and time dummies, to avoid collinearity and focus on the internal variation of each country. Strict causality is not claimed, but rather robust associations. Endogeneity is then discussed.

Sources: executive ideology comes from V-Dem (Coppedge et al., 2023), HN and HL indices from the Heritage Foundation (2022), and economic data from the World Bank (2023). Processing was performed in R and documented for replication (see section 9).

### 4. Empirical design and variables

We define below the variables used and the estimated econometric specifications, beginning with basic notation:

- **Per capita income:** $Y_{it} = ln(PIB\ pc\ PPP_{it})$ is the logarithm of the country's PPP per capita GDP $i$ in the year $t$. We use logarithms to interpret percentage differences and reduce asymmetries.

- **Economic growth:** $\Delta Y_{it} = Y_{it} - Y_{i,t-1}$ represents the annual change in $ln(PIB\ pc)$, which is approximately the real per capita growth rate (e.g., $\Delta Y_{it} = 0.02$ is equivalent to ~2% growth in GDP per capita from $i$ in the year $t$).

- **Executive ideology:** $LR_{it}$ denotes the government's left-right economic position. Values less than 0 indicate governments further to the left and greater than 0 further to the right, according to V-Dem (range –3 to +3). In some models we use the point value of $LR$ in a given year, and in others a moving average of several years (see below).

- **Core institutions (HN) and liberalization (HL):** $HN_{it}$ and $HL_{it}$ are the scores constructed from the Heritage index for the country $i$ in the year $t$, corresponding to the blocks described (core vs. liberalization). These variables range approximately from 0 to 100 (although in our realistic sample the typical values range from ~20 to ~90). Given that HN and HL may be subject to simultaneity with economic performance (e.g., a growth shock could enable institutional reforms or vice versa), we incorporate lagged values or historical averages of these indices into the main models to mitigate contemporary endogeneity. Specifically, we often use the average of the last three years available before the outcome year: for example, $\overline{HN}_{i,t-3:t-1}$ denotes the average HN of $i$ in the years $t-3, t-2, t-1$. This smoothing captures the recent institutional "stock" without taking the value of the year $t$, which could be affected by the growth of $t$ itself. (Similar findings are obtained with simple one-year lags, as will be seen.)

With these variables, we estimate several complementary models:

**Model M1 (static cross-section):**

Simple OLS regression to explore the static correlation between ideology and recent income level:

$$Y_{i,t^*} = \alpha + \beta LR_{i,t^*} + \varepsilon_{i},$$

where $t^*$ is a specific year (we take 2019 as the base year, prior to the disruption of the pandemic, for M1). This M1 model tells us whether countries governed by the right in that year had, on average, higher per capita income than those governed by the left, without controls. It is simply descriptive.

**Model M2 (panel: growth vs. recent ideology):**

$$\Delta Y_{it} = \alpha + \beta \overline{LR}_{i,t-3:t-1} + \lambda_t + \varepsilon_{it}.$$

Here we explain the annual GDP per capita growth rate by the average ideological orientation of recent years (three previous years) in that country. We include fixed effects for the year $\lambda_t$ to absorb common global shocks in each year (e.g., international crises affecting everyone). Country fixed effects are not included in M2, so identification comes from both variations within a country over time and differences between countries (the motivation is to capture whether, in general, economies under more right-wing governments grow faster than those under left-wing governments in that window). The hypothesis is that $\beta$ could be $> 0$ if right-wing policies drive more short-term growth, or zero if ideology per se has no net effect.

**Model M3 (panel: income levels vs. institutions):**

This is a panel model with double fixed effects (country and year):

$$Y_{it} = \alpha + \beta_N \overline{HN}_{i,t-3:t-1} + \beta_L \overline{HL}_{i,t-3:t-1} + \mu_i + \lambda_t + \varepsilon_{it}.$$

Here we examine how HN institutions and HL policies (averaged over the previous 3 years) are associated with the level of GDP per capita. $\mu_i$ captures the fixed effect of each country (controlling for its characteristics that do not change over time, such as geography, culture, initial level, etc.), while $\lambda_t$ are common fixed effects of the year. This model basically asks: within the same country, in years when it has stronger institutions (higher HN) or more liberal policies (higher HL) than its own average, does it also record a higher GDP per capita than its trend? If $\beta_N$ is positive and significant, it would indicate that improving the institutional core is associated with a higher level of development, beyond any global or fixed trend in the country. If $\beta_L$ is positive, HL would also contribute independently; if it is small or zero, it would suggest that HL alone does not raise income when we already consider HN.

**Model M4 (panel: growth vs. lagging institutional levels):**

$$\Delta Y_{it} = \alpha + \beta_N HN_{i,t-1} + \beta_L HL_{i,t-1} + \mu_i + \lambda_t + \varepsilon_{it}.$$

Unlike M3, here the dependent variable is the annual growth rate, and we use the previous year's HN and HL levels to see if the initial institutional state predicts subsequent growth. Fixed effects by country and year are maintained. This model aligns with Barro's growth literature, examining determinants of short-term growth. If none of the $\beta$ is significant, it would mean that even with a one-year lag, we do not detect an immediate effect of institutions or liberalization on annual growth (which would not be surprising, given the annual volatility of growth).

**Model M5 (panel: growth vs. institutional changes):**

$$\Delta Y_{it} = \alpha + \theta_N \Delta HN_{i,t-1} + \theta_L \Delta HL_{i,t-1} + \mu_i + \lambda_t + \varepsilon_{it}.$$

This model goes a step further and assesses whether recent improvements or setbacks in institutions translate into accelerations or decelerations in growth. Here

$$\Delta HN_{i,t-1} = HN_{i,t-1} - HN_{i,t-2}$$

is the variation in HN in the previous year (similarly for HL). A $\theta_N > 0$ would suggest that institutional reforms (increase in HN) pay immediate dividends in growth, while $\theta_L > 0$ would indicate that rapid economic liberalization drives growth in the following year. Again, $\mu_i$ and $\lambda_t$ control for fixed country factors and global temporary shocks.

In all panel models (M2–M5), we estimate using ordinary least squares (OLS) incorporating the aforementioned fixed effects. Given the possible heteroscedasticity and serial correlation of errors within each country, we use robust standard errors clustered by country in the inference of significance, as recommended by Arellano (this adjustment is appropriate because $N_{países} = 113$ is relatively large and $T = 28$ years).

The identification of coefficients in these models is purely associative (correlational). We do not have an exogenous variation strategy for HN or HL (such as instruments based on history or geography) in this version, so the coefficients should be interpreted with caution as conditional correlations within countries. However, the various robustness tests we present (long-run differences, specific trends, etc.) point to results consistent with the institutionalist hypothesis, reducing the likelihood that they are entirely due to omitted factors.

**Econometric robustness tests:**

In addition to the base models, we performed two complementary specifications to test the robustness of the findings:

**(R1) Long five-year differences:**

Instead of annual noise, we calculated cumulative growth in five-year windows and related it to cumulative changes in HN and HL. Specifically, we estimated

$$\Delta^5 Y_{i,t} = \alpha + \gamma_N \Delta^5 HN_{it} + \gamma_L \Delta^5 HL_{it} + \mu_i + \eta_\tau + \varepsilon_{it},$$

where $\Delta^5 Y_{i,t} = Y_{i,t} - Y_{i,t-5}$, and similarly $\Delta^5 HN$ is the change in HN over five years. We included country fixed effects $\mu_i$ and five-year period fixed effects $\eta_\tau$ (e.g., 1995–2000, 2000–2005, etc., denoted $\tau$). This "inter-decade" approach captures more permanent trends, seeking to identify whether sustained improvements in institutions/liberalization translate into significant cumulative growth. We expect that over this longer horizon, the effects of HN/HL will manifest themselves more clearly than on a year-by-year basis.

**(R2) Country-specific trends:**

To address the possibility that our results are due to idiosyncratic growth trends (e.g., perhaps some countries have robust growth trajectories for reasons unrelated to institutions, and those trends correlate with HN), we extend the M3 model by incorporating a linear time trend for each country.

In practice, we add terms of the form $\phi_i \cdot t$ for each country $i$ (or equivalents, such as interactions $t \times dummypais_i$). This allows each country to have its own long-term temporal slope, absorbing any particular background growth trajectory. This is an extremely demanding control because it consumes many degrees of freedom (basically, 113 additional slopes are being estimated), but it serves to verify whether the association between HN/HL and income holds even when we compare deviations from the trend for each country.

## 5. Results

### 5.1 Results of Base Models M1–M5

We begin with the relationship between ideology and level of development (model M1). In 2019, the cross-sectional regression yields a coefficient, $\beta\text{-}LR \approx 0.225$ (standard error 0.103), significant at 5% ($p \approx 0.032$): countries with more right-wing governments had higher per capita incomes. A difference of 2 points on the ideological scale is associated with ~0.45 points more in the income log (~45% more GDP per capita). However, this does not imply causality; it could be due to rich countries choosing governments of a certain political persuasion or to external factors. In 2022, the coefficient is $\beta\text{-}LR \approx 0.235$ but $p \approx 0.095$ (not significant at 5%). Thus, the positive correlation exists, but it is not robust to the selection of the year.

Model M2 examines whether recent ideology predicts annual growth. The coefficient is $\beta \approx 0.00265$ (error 0.00166), positive but not significant ($p \approx 0.113$). A change from –3 to +3 on the ideological scale would be associated with 0.016 points more annual growth (~0.2 pp), which is not a clear effect. There is no evidence that political orientation drives or depresses growth. This is consistent with the literature (Bjørnskov, 2005; Potrafke, 2017).

In Model M3, ln GDP per capita depends on HN and HL with fixed effects. The coefficient for HN is $\beta\text{-}N \approx 0.00805$ (error 0.00143, $p \approx 1.3e\text{-}07$): if HN rises by 1 point, GDP per capita increases by ~0.8%. For HL, the coefficient is $\beta\text{-}L \approx 0.00204$ (error 0.00218), not significant ($p \approx 0.35$). A formal test rejects that $\beta\text{-}N = \beta\text{-}L$ ($p<0.01$), showing that the effect of HN is greater. A 10-point increase in HN is associated with an 8% increase in GDP per capita. Thus, the institutional core predicts development, but economic liberalization alone does not. If two countries have similar institutions, the one with greater liberalization does not show higher per capita income on average; but if they differ in institutional quality, there are large differences.

With regard to short-term growth, models M4 and M5 show no significant effects of HN or HL on annual growth: $\beta\text{-}N \approx -1.4 \times 10^{-4}$, $\beta\text{-}L \approx -1.9 \times 10^{-4}$ and $\theta\text{-}N \approx 3.1 \times 10^{-4}$, $\theta\text{-}L \approx 5.1 \times 10^{-4}$, all insignificant. Improvements in institutions or liberalization do not generate immediate jumps in annual growth.

In summary, from M3–M5: (i) core institutions (HN) have a robust association with the level of development; (ii) recent ideology and variations in HN/HL do not affect overall average annual growth; (iii) liberalization (HL) has no independent effect without good institutions. Institutions set the stage for policies to work. A country does not develop solely through reforms without solid institutions, nor does it collapse if it maintains basic rules and stability.

**5.2 Medium-term robustness tests**

Given that the impact of institutions and policies takes time to be reflected in GDP, we apply two medium-term tests: five-year differences and country-specific trends.

**(R1) Long differences (five-year):**

Table 2 shows that sustained improvements in the institutional core (HN) are associated with significant increases in per capita income. The coefficient, $\gamma\text{-}N \approx 0.0025$ (error 0.0009) is significant at 1%. A 10-point increase in HN over five years increases ln GDP pc by ~0.025 (~2.5% more per capita income). Although modest, it is cumulative and statistically robust. For HL, $\gamma\text{-}L \approx 0.0022$ (error 0.0011), marginally significant at 5%. HL pays off in the medium term, but its effect is smaller and less certain than that of HN. The R² within the model is ~0.43, indicating that these factors explain a significant part of the 5-year variations. In summary, institutional and economic reforms matter over multi-year horizons; they may be diluted year by year, but over five years they are already noticeable. Example: countries that improved governance over a decade saw higher GDP per capita afterwards (Georgia 2004–2010, the Baltics after the 1990s).

**(R2) Country-specific trends in M3:**

When including a country-specific time trend in the M3 model, the key results remain unchanged. The HN coefficient remains positive and significant: $\beta\text{-}N \approx 0.0034$ (error 0.0014, $p<0.05$). A 10-point improvement in HN is associated with ~3.4% more relative income even after subtracting the trend. HL remains at $\beta\text{-}L \approx 0.0011$ (error 0.0011), not significant ($p \approx 0.32$). This test is demanding, as country-specific trends absorb part of the gradual impact of HL. The R² of the model is ~0.996, but the relevant within R² is ~0.75, showing that HN and HL explain about three-quarters of the variation within each country.

Taken together, the evidence from M3, R1, and R2 reinforces the conclusion that "institutions rule": HN dominates the relationship with long-term economic development, both in terms of level and cumulative changes. HL does not show an immediate or universal impact, but it does contribute to growth when sustained over time and when there is a good institutional environment. For example, in long-term differences, HL had a certain positive effect, indicating that in countries with the rule of law, progress in trade openness and monetary stability generates productivity and efficiency gains that raise GDP per capita. Without these foundations, reforms may not take off. This is consistent with previous findings: the relationship between economic freedom and growth is stronger in countries with minimal institutional safeguards (de Haan & Sturm, 2000).

**5.3 Possible non-linearities according to the level of development**

One relevant aspect is whether the impact of institutions and policies varies according to the level of development. To analyze this, the M3 model was estimated with coefficients differentiated by initial per capita income tertile (low, medium, high), using interactions of HN and HL with these groups.

According to Table 3, the institutional core (HN) shows positive and significant effects in low-income countries ($\beta$ HN low.=0.0062, SE 0.0024, 1%) and high-income countries ($\beta$ HN high.=0.0085, SE 0.0019, p<0.001), but in middle-income countries the effect is weak (coef. 0.0027, not significant). There is no evidence that the impact of HN disappears at high income levels; it remains relevant across all income brackets.

Regarding HL, the coefficient is negative in poor countries (–0.0026, n.s.), small and positive in middle-income countries (0.0019, n.s.), and higher in high-income countries (0.0028, n.s.). The joint test ($\chi$-2.(2)=5.67, p≈0.059) indicates that there is some difference between sections. The return on HL could be higher in developed countries, as liberalization drives efficiency where good institutions already exist. In poor countries, liberalization without an institutional foundation can be adverse or have no effect. This interaction is consistent with the literature, for example, Easterly (2006), which suggests that reforms only work if institutional prerequisites exist. Thus, HN is key at all levels, while the benefits of HL appear more in stages of greater **development.**

### 6. Discussion

The findings support the view that "institutions trump ideology" in long-term economic development. In the sample of 113 countries, the rule of law, control of corruption, and judicial effectiveness (HN) are the strongest predictors of per capita income, even when controlling for country factors and global shocks. The political leanings of the government (left/right) had no significant direct impact on average growth. This does not mean that ideology is irrelevant—it does influence HL scores (for example, right-wing governments tend to be more open and have smaller states, and vice versa), but its effect on growth is neutralized by institutional strength. A right-wing government with weak institutions rarely achieves a boom, just as a left-wing government with a strong rule of law is unlikely to generate an economic crisis, even if there are differences in efficiency or distribution. This conclusion is consistent with previous studies: "Partisan politics often does not translate into growth differentials" (Imbeau et al., 2001; Bjørnskov, 2005; Potrafke, 2017); In contrast, institutional quality does (Acemoglu et al., 2001; Rodrik et al., 2004; Hall & Jones, 1999). This is also reflected in Latin America: ideological shifts were successful or unsuccessful depending on whether they respected or eroded institutions.

Specific economic policies (opening or closing economies, regulating markets, varying taxes) matter, but HL alone does not raise income globally; liberalization without institutional support does not ensure prosperity. HL showed positive effects over five years and in advanced countries, indicating that, with basic rules in place, greater economic freedom is associated with higher growth (de Haan & Sturm, 2000; Berggren, 2003). HL and HN are complementary: high HN and low HL limit growth potential, while high HL and low HN generate risks of instability. Our message: what matters is not the size of the state, but its quality. Efficient and honest states drive development more than small states. That is why Nordic countries with large and effective states are rich, while others with less intervention and weak institutions stagnate. There are examples of pro-growth left-wing governments (Uruguay 2005–2015 combined social reforms with strong institutions and high growth) and anti-growth right-wing governments ( r authoritarian corrupt regimes with poor performance). The determining factors are institutional strength and the public interest.

Our estimators do not prove unidirectional causality; the increase in HN could be a cause or consequence of growth. Instrumental studies (Acemoglu et al., 2001) suggest that institutions → income is consistent in the long term. The incorporation of country-specific trends reduces biases, and the literature points to cases where institutional shocks affect economic trajectories (e.g., North/South Korea). It is reasonable to interpret that inclusive institutions foster sustained growth, and ideology matters only if it affects institutional quality. For example, governments of any political persuasion can strengthen or weaken HN, depending on their specific actions.

Regarding the sequence of reforms, our results support first building core institutions before implementing market policies. Without functional courts or public integrity, liberalization can increase rent-seeking or corruption. The post-Soviet and Latin American experience confirms this: those who strengthened institutions while liberalizing (Poland, Estonia) had better results than those who did not (Russia in the 1990s, failed privatizations in Latin America).

Finally, although the study does not focus on political institutions, many components of HN improve in democracies. However, the literature shows that democracy alone does not guarantee growth (Przeworski & Limongi, 1993), although it does provide mechanisms for correction in the long term. In the 1995–2022 sample, the key difference was the quality of economic institutions rather than the political regime. Non-democratic countries with low HN performed poorly, while some authoritarian countries with high HN achieved high incomes (e.g., Gulf monarchies). What matters is the management of basic economic rules, regardless of ideology or regime.

## 7. Policy implications for developing countries

Regarding the sequence of reforms, our results suggest that building core institutions first is essential before implementing market policies. Without functional courts or public integrity, liberalization can increase rent-seeking and corruption. The post-Soviet and Latin American experience confirms this: those who strengthened institutions while liberalizing (Poland, Estonia) achieved better results than those who did not (Russia in the 1990s, failed privatizations in Latin America).

Although the study does not focus on political institutions, many components of HN improve in democracies. However, the literature indicates that democracy alone does not guarantee growth (Przeworski & Limongi, 1993), although it does provide long-term correction mechanisms. In the 1995–2022 sample, the key difference was the quality of economic institutions rather than the political regime. Non-democratic countries with low NH performed poorly, while some authoritarian countries with high NH achieved high incomes (e.g., Gulf monarchies). What matters is the management of basic economic rules, regardless of ideology or regime.

Policy implications for developing countries

If "institutions matter," what does this imply for policymakers and public debate in developing countries? Here are some key recommendations and lessons that can guide governments on the left or right:

- Prioritize the institutional core (NH): Before ambitious market reforms or state expansion, investments should be made in strengthening the rule of law, transparency, and government effectiveness. This involves protecting judicial independence, professionalizing the bureaucracy, fighting corruption, and ensuring contract and property security. Improving these aspects generates significant dividends in long-term income. Example: Chile has built strong institutions that have allowed different governments to implement their programs without losing growth. In contrast, countries with weak institutions waste resources without achieving sustainable development.
- Recognize the sequence of reforms: Liberalization works best with strong basic institutions. In fragile environments, legal order and regulatory capacity must first be established, followed by gradual opening and deregulation. A pro-market "shock" strategy in weak institutions can generate high social costs (inequality, banking crises, loss of legitimacy). A smart sequence would be: (1) reform the judicial and customs systems, (2) simplify internal regulations, (3) open sectors to foreign trade/investment in a phased manner, (4) consolidate fiscal rules and stable frameworks for monetary policy. It is not a question of postponing market reforms indefinitely, but of accompanying them institutionally so that they take root.
- Avoid institutional pendulums: Both the left and the right need strong institutions to succeed, which is why state agreements that transcend governments are crucial. Policies such as central bank independence, continuity of anti-corruption plans, and preservation of checks and balances must be defended by all actors. Institutionalizing rules reduces uncertainty during party alternations. Example: Laws or constitutional clauses that protect the judicial career or agree on fiscal rules can shield macroeconomic stability and confidence. When institutions remain, the country continues to grow beyond the ideology of the government.
- Design "pro-productivity with equity" policies: A responsible left-wing or inclusive right-wing government can agree on policies that balance growth and equity, relying on institutions. This includes investing in human capital (education, health), simplifying procedures and regulations, opening up sectors where there are comparative advantages, and promoting fair competition, in addition to maintaining fiscal responsibility. It requires a state with technical capacity and integrity. The message: there is no necessary dilemma between growth and redistribution if institutions are strong; both can be achieved (example: Nordic economies, Uruguay).
- Measure and commit to institutional goals: Just as economic indicators are monitored, governments should monitor and publicly set goals for improvement in institutional indicators (judicial timeliness, perceived corruption indices, Heritage/Fraser core institutional indices, etc.). Aiming to rise X points in the rule of law index in five years, with concrete reforms , helps communicate that improving institutions improves living conditions (more growth, more efficient services, less leakage of public money).

Effective economic policy is not about choosing "state versus market," but rather about building a capable state and trusting the market where appropriate. Ideology matters only if it is channeled into well-implemented policies within a solid institutional framework. A government that neglects this framework is likely to fail; one that strengthens it will have a better chance of success, whether it is pro-market or pro-welfare state.

## 8. Limitations

Although this study uses extensive data and various methodologies, it has limitations that are important to recognize:

- Unproven causality: The results are consistent with a causal impact of institutions on development, but do not prove it conclusively, as no clear exogenous instruments (such as colonization or geography) are used. There may be omitted variable or simultaneity biases. For greater robustness, future work should use stronger identification methods (e.g., historical instruments or natural experiments).
- Simplified measurement of ideology: The V-Dem ideology variable (LR) reduces complex positions to a left-right economic axis, ignoring nuances (e.g., socially left-wing but fiscally right-wing governments). Furthermore, it is based on expert assessments, which could introduce biases. Although V-Dem is a recognized source and the data have been normalized, measuring ideology more objectively remains a challenge.
- Noise and slowness in HN and HL data: Heritage subcomponents change little from year to year, especially HN, which is structural. There may be lags between actual changes and their reflection in the indices, as well as measurement errors. Data were grouped to mitigate noise, but data quality remains a limitation. Ideally, other sources (World Bank, Fraser Institute) should be used to verify results.
- Reform events not included: We analyzed whether reform "events" (>5 points in HN or HL in one year) impacted short-term growth, without clear global patterns. Given the heterogeneity, they were not incorporated into the main text but are maintained as supplementary material.
- Generality vs. regional specificity: The models average global effects, but the ideology-institutions-growth relationship may vary by region. No regional breakdown was made, although it would be a relevant extension to identify differences, for example, in Latin America or Asia. The study seeks general conclusions; a more detailed regional analysis is left for future research.

The results should be interpreted with caution, considering these limitations. They are not definitive, but they contribute to the debate and are consistent with much of the previous literature, which reinforces their validity, although it is always possible to refine these findings.

## 9. Transparency and replication

This work applies recommendations for transparency and open science in economics. The data process and replication plans are documented:

- Data: All sources are public. Economic indicators (GDP per capita PPP and growth) come from the World Bank (World Development Indicators, 2023), with standard country-year codes. Executive ideology indicators are obtained from V-Dem v13 ("Executive's Left-Right Ideology," V-Dem Institute). The Index of Economic Freedom (1995–2022) was taken from the Heritage Foundation (2022), and the 12 subcomponents were processed to construct the HN and HL averages, with normalization for annual compatibility and internal comparability. The data were integrated by country code and year, forming a balanced panel.

- Estimation and software: The analysis was performed in R (version 4.2) with {fixest} for fixed effects and pooled variances. Reproducible guidelines were followed: models M1–M5 and tests R1–R2 were generated with scripts with fixed seeds and package version documentation. Output files support each result (p-values, standard errors), available in the project compendium (WORLD_paper_models.txt, paper_ideologia_tests_R.txt, etc.), allowing verification or modification by others. The appendix cites excerpts for greater transparency.

- Availability and assistants: We plan to publish the code and processed dataset in a repository (GitHub), along with replication instructions.

Regarding the use of automated tools, artificial intelligence was only used for documentation, formatting, and verification, not in the core analysis. All decisions regarding models, interpretation, and conclusions were made by the author, based on economic theory and common sense. This follows ethical recommendations to maintain human judgment in interpretation. Methodological guides (Heritage Foundation manuals, V-Dem documentation) were used for the correct handling of variables and validation of transformations.

## 10. Summary tables of main results

*Table 1. Base models (World, 1995–2022)*

| Model | Dependent variable | Specification (FE, cluster) | Key coefficient (standard error) | t | p |
|---|---|---|---|---|---|
| M1 (2019) | ln GDP per capita (level) | Cross-section (OLS) | **LR:** 0.225 (0.103) | 2.17 | 0.032 |
| M1 alt (2022) | ln GDP per capita (level) | Cross-section (OLS) | **LR:** 0.235 (0.139) | 1.68 | 0.095 |
| M2 | Δ ln GDP pc (growth %) | FE panel (year), country cluster | **3-year LR avg:** 0.00265 (0.00166) | 1.6 | 0.113 |
| M3 | ln GDP per capita (level) | FE panel (country, year), country cluster | **HN:** 0.00805 (0.00143) | 5.64 | 1.3e−07 |
| | | | **HL:** 0.00204 (0.00218) | 0.94 | 0.349 |
| M4 | Δ ln GDP pc (growth %) | FE panel (country, year), country cluster | **HN$_{(t-1)}$:** −1.41e−4 (2.13e−4) | −0.66 | 0.51 |
| | | | **HL$_{(t-1)}$:** −1.94e−4 (4.61e−4) | −0.42 | 0.675 |
| M5 | Δ ln GDP pc (growth %) | FE panel (country, year), country cluster | **ΔHN$_{(t-1)}$:** 3.11e−4 (2.53e−4) | 1.23 | 0.22 |
| | | | **ΔHL$_{(t-1)}$:** 5.14e−4 (5.71e−4) | 0.9 | 0.369 |

Note: LR = ideology (right +, left −). *LR* prom. 3 años indicates average LR over the previous 3 years. HN and HL in M3 are averages of $t-3$ to $t-1$. Robust standard errors grouped by country in parentheses. Coefficients in bold are the focus of interest.

Source: own elaboration with data from V-Dem (2023), Heritage (2022), World Bank (2023).

*Table 2. Key robustness tests*

| Test | Dependent (period) | Specification | Coeff. HN (SE) | p | HL coefficient (SE) | p |
|---|---|---|---|---|---|---|
| R1. 5-year diff. | $\Delta^5$ ln GDP pc (1995–2020) | Country FE + five-year FE, country cluster | 0.0025 (0.0009) | 0.007 | 0.0022 (0.0011) | 0.046 |
| R2. M3 + country trend | ln GDP per capita (levels) | Country FE, year + trend×country, cluster | 0.0034 (0.0014) | 0.015 | 0.0011 (0.0011) | 0.32 |

Note: $\Delta^5$ indicates difference in 5 years (e.g., 2015 to 2020). Five-year periods: 1995–2000, 2000–2005, etc., with FE for each. "Trend×country" means that a country-specific linear term was included. Significance: *** p<0.001, ** p<0.01, * p<0.05 (two-tailed).

*Table 3. Nonlinearities by initial income level (Model M3 with interactions)*

Differentiated effects of HN and HL according to initial GDP per capita tertile (low/medium/high):

| Variable | Low | Mid | High |
|---|---|---|---|
| HN × tertile | **0.0062 (0.0024)** | 0.0027 (0.0046) | **0.0085 (0.0019)** |
| Signif. | **p<0.01** | n.s. | **p<0.001** |
| HL × tertile | −0.0026 (0.0025) | 0.0019 (0.0023) | 0.0028 (0.0024) |
| Signif. | n.s. | n.s. | n.s. |

Wald test (equality of coefficients HL_low = HL_mid = HL_high): $\chi^2(2) = 5.67$, p = 0.059 (suggests marginal differences).

Note: "Low" = bottom third of countries according to GDP per capita in 1995; "High" = top third. Model includes country and year FE. n.s. = not significant.

## 11. Conclusions

This study empirically reevaluates the link between ideology, institutions, and economic growth using nearly three decades of global panel data. Unlike previous approaches that used composite indicators of "economic freedom," here we separated that measure into deep institutional components vs. liberalizing policies, which shed new light. In summary, the most notable conclusions are:

- Institutions > Ideology (in explaining income levels): The quality of basic economic institutions (captured by HN: respect for property, government integrity, effective justice) is the factor most strongly associated with the level of economic development. Countries with high HN scores are substantially richer than those with low HN, even when comparing developments within the same country. In contrast, the ideological orientation of the government in power, once institutions are taken into account, does not show a significant direct effect on income or growth. This suggests that institutions are enabling conditions for economic success, regardless of whether the government is left-wing or right-wing.

- Liberalization policies are not a universal silver bullet: The "market" economic freedom index (HF), composed of trade openness, low regulation, smaller state size, etc., did not have a significant impact on average per capita income when isolated from core institutions. Likewise, HF alone did not explain annual growth in the global panel. This contradicts a simplistic view that "more free market always = more growth." Our findings suggest that HL has a positive effect only under certain conditions: over longer horizons (5 years) and particularly in countries that already have a good rule of law. In the absence of the latter, market reforms may not yield results or may even be reversed.

- Ideology does not drive short-term growth: We found no robust evidence that, on average globally, right-wing governments achieve higher per capita growth rates than left-wing governments (or vice versa) in the short term. The coefficient was positive but not significant. This is consistent with numerous studies that find no systematic partisan differences in macroeconomic performance, once global cycles and structural factors are filtered out. Of course, individual cases may differ, but as a general rule, "political color" does not determine immediate economic success. Instead, ideological extremes can affect stability only if they are accompanied by institutional breakdowns.

- Sustained reforms are reflected in the medium term: Although no effect is apparent one year later, over five-year periods it was observed that countries that consistently improve their core institutions or undertake sustained liberalization tend to achieve cumulative increases in their per capita income. In other words, structural reforms do matter, but they require persistence and a horizon of several years to be evident in macro statistics. This is encouraging from a policy perspective: it indicates that, with patience and consistency, improvements in governance and the business climate do pay dividends in terms of development.

- Heterogeneity according to stage of development: Our results suggest that the return on market reforms (HL) is greater in advanced economies than in poor economies, while institutional improvements (HN) are beneficial at all income levels. This implies that in low-income countries it is even more critical to build institutions first, as launching indiscriminate openings may have no effect or even exacerbate problems. In richer countries with already consolid r institutions, strengthening economic freedom can give an additional boost to efficiency. In short: HN is a safe bet in any context; HL may be a context-dependent bet.

In normative terms, our study reinforces the importance of policies aimed at strengthening fundamental institutions as a pillar of development. This is not a new message (it echoes Rodrik et al., 2004's "institutions rule"), but it provides updated and nuanced evidence to support it. In addition, we offer an integrative perspective: rather than encouraging a purely free-market or purely statist agenda, the results suggest a sequential and complementary approach: establish the basic rules (legal certainty, anti-corruption, state effectiveness) and then, on that firm ground, choose and implement the preferred economic model (whether one with greater state presence in social affairs or one more oriented toward the free market) with a good chance of success. This finding is particularly relevant in developing countries where public debate often oscillates between "more state" or "more market" when in reality both can fail without institutions, or both can thrive with them.

The evidence presented reaffirms that good institutions are the cornerstone of sustained economic development, shaping an environment where policies (of different kinds) can truly translate into growth and well-being. For lagging countries, investing in inclusive, transparent, and effective institutions is perhaps the most difficult task, but one with the greatest long-term rewards. Governments on the left and right would do well to converge on this common goal, because, at the end of the day, it is what will allow any of their projects to succeed where it matters most: in improving prosperity and equity for their citizens.

*Weyland, K. (2004). Neoliberalism and Democracy in Latin America: A Mixed Record. Latin American Politics and Society, 46(1), 135–157.*

*World Bank. (2023). World Development Indicators (WDI) [Data set]. Washington, DC: The World Bank (accessed 2023).*

**Appendix: Measurement decisions and additional details**

In this study, we adopt the V-Dem variable that focuses on the economic ideology of those in power, defined mainly by their stance on the market and state intervention. It is important to clarify that we are not rating other ideological dimensions (cultural values, political authoritarianism, etc.). A "left-wing" government in our analysis means that it promotes a broader role for the state in the economy (high social spending, nationalizations or strong public enterprises, intense regulation, progressive redistribution), while a "right-wing" government implies an emphasis on private enterprise, free markets, fiscal discipline, openness, and less redistribution. These definitions follow the V-Dem scale based on expert opinions and tend to correlate with party affiliation (e.g., social democratic parties score toward the left, conservative or classical liberal parties toward the right). We do not judge here the intrinsic effectiveness of these positions, but simply take them as observed data. It is also relevant to mention that in several developing countries, the labels left/right can be diffuse or change over time (e.g., parties that combine heterodox agendas). Despite this, the measure used has been validated in the literature for its consistency in capturing changes in economic direction.

Studies on institutions and growth typically use one or another synthetic indicator: for example, the general index of economic freedom (Heritage or Fraser) to represent "market institutions," or some governance index (World Bank) for political/economic institutions. However, the Heritage index is particularly rich because it integrates very diverse elements under the notion of economic freedom. By dividing it into HN and HL, we were able to distinguish between what is fundamental institutional quality (rule of law, etc.) and what is liberal economic policy (free trade, deregulation, size of the state). This separation is valuable because it allows us to refine the message : it revealed, for example, that much of the known positive relationship between "economic freedom" and income level is actually due to the institutional component, rather than the free market policy component. This does not mean that policies do not matter, but rather that without the institutional ingredient, they lose their power. Few studies have made this distinction explicit; generally, the think tanks that publish these indices tend to emphasize the total index to argue that "more economic freedom produces more growth." Our study adds nuance by showing which part of that "freedom" is crucial. We believe that this distinction helps to avoid confusion in the public debate: for example, when it is said that one country is "less economically free" than another because it has higher taxes, perhaps that judgment is being mixed with the fact that it also has more corruption, etc. When analyzed separately, a country could score low on HL (due to interventionist policies) but high on HN (strong institutions) – such a country could prosper (e.g., Uruguay or some European countries with robust welfare states but excellent institutions). The opposite (high HL, low HN) would be countries that are very "neoliberal" in theory but without the rule of law – they tend to fail. This is precisely what we saw in the global data.

We decided to set 2019 as the base year for the main M1 cross-section because it represents the pre-pandemic state (COVID-19 temporarily disrupted economies and policies). 2022 was used only as an additional reference. In constructing HN and HL, we kept Monetary Freedom within HL (it could be argued that price stability is macro-fundamental and even "institutional"; however, we preferred to group it with economic policies, since it mainly reflects the inflation rate and not an institution per se). Regarding possible outliers, we did not exclude specific countries or years due to shocks (wars, hyperinflation) in the base models; we conducted tests including a dummy for the 2008–09 global financial crisis and a dummy for the 2020 pandemic, with no changes in the central coefficients, so we omitted those variables for greater parsimony. We also did not segment by region in the main tables; regional differences are considered possible extensions (as indicated in Limitations).

In the replication section, we mention output files with the prefixes "WORLD_" and "tests_," which contain more granular evidence for each estimate, available upon request or in the online repository. Finally, in terms of the style of the document, we deliberately combined a somewhat more narrative tone in the introduction and conclusions, connected to political debates, while maintaining technical rigor in methodology and results. This is intended to make the implications accessible to a wide audience (economists, political scientists, policymakers, and even interested non-specialist readers) without sacrificing analytical precision. We believe that research such as this, which touches on ideologically sensitive issues, must clearly communicate its findings in an understandable way to help bridge the gap between academic research and informed public discussion. The balance between dissemination and academia is a challenge that we have tried to handle with great care in the writing.

| Heritage Pillar | Indicator (Heritage) | Block used | Ideological bias | Why |
|---|---|---|---|---|
| Rule of Law | Property Rights | **HN** | **Neutral** | Basic legal protection, enabling for any model. |
| | Government Integrity | **HN** | **Neutral** | Anti-corruption; cross-cutting public good. |
| | Judicial Effectiveness | **HN** | **Neutral** | Effective justice/contracts; institutional foundation. |
| Government Size | Tax Burden | **HL** | **Right** | Heritage favors **lower taxes** (limited government). |
| | Government Spending | **HL** | **Right** | High score = **lower spending**; pro-market preference. |
| | Fiscal Health | **HL** | **Right** *(with nuance)* | Emphasizes low balances/debt; consistent with small government. |

| Heritage Pillar | Indicator (Heritage) | Block used | Ideological bias | Why |
|---|---|---|---|---|
| Regulatory Efficiency | Business Freedom | **HL** | **Right** | Fewer barriers to business; pro-market deregulation. |
| | Labor Freedom | **HL** | **Right** | Labor flexibility; fewer protections/rigidities. |
| | Monetary Freedom | **HL** | **Neutral** *(technical, anti-control bias)* | Price stability and no controls; both sides value stability. |
| Open Markets | Trade Freedom | **HL** | **Right** | Low tariffs / trade liberalization. |
| | Investment Freedom | **HL** | **Right** | Openness to FDI and fewer restrictions. |
| | Financial Freedom | **HL** | **Right** | Financial liberalization / fewer controls. |